\documentclass[preprint,authoryear,12pt]{elsarticle}

\usepackage{epsfig}

\usepackage{amssymb}

\usepackage[ps2pdf,%
a4paper=true,%
breaklinks=true,%
colorlinks=true,%
pdfauthor={First Author et al.},%
pdftitle={Template for manuscripts in Advances in Space Research}%
]{hyperref}

\journal{Advances in Space Research}

\begin{document}

\begin{frontmatter}



\title{Cosmic-ray physics with IceCube}


\author{Thomas K. Gaisser}
\address{Bartol Research Institute and Dept. of Physics and Astronomy\\
University of Delaware, Newark, DE, USA}
\ead{gaisser@bartol.udel.edu}


\author{for the IceCube Collaboration}
\address{http://www.icecube.wisc.edu/collaboration/authorlists}


\begin{abstract}

IceCube as a three-dimensional air-shower array covers an energy range of the cosmic-ray spectrum from below
1 PeV to approximately 1 EeV.  This talk is a brief review of the function and goals of IceTop, the surface component
of the IceCube neutrino telescope.  An overview of different and complementary ways that IceCube is
sensitive to the primary cosmic-ray composition up to the EeV range is presented.  Plans to
obtain composition information in the threshold region of the detector in order to overlap with direct
measurements of the primary composition in the 100 -- 300 TeV range are also described.

\end{abstract}


\end{frontmatter}

\parindent=0.5 cm

\section{Introduction}

The principal goal of IceCube is to identify and measure high-energy neutrinos from extra-terrestrial sources,
thus opening a new view into acceleration of protons and nuclei in energetic cosmic sources.  
The recently completed IceCube neutrino telescope was
described in a separate talk in this session by~\citet{Ty},
which gives a status report on searches for astrophysical neutrinos with
the partially completed detector.  

\begin{figure}
\begin{center}
\includegraphics*[width=9.5cm]{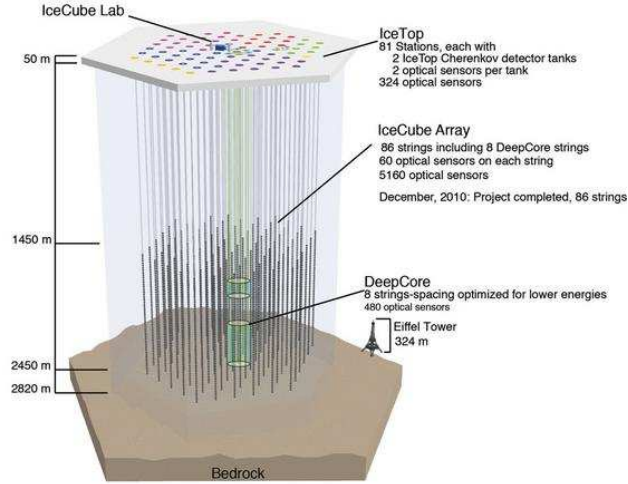}
\end{center}
\caption{Configuration of IceCube.}
\label{fig0}
\end{figure}

\begin{figure}[thb]
\begin{center}
\includegraphics*[width=4.5cm]{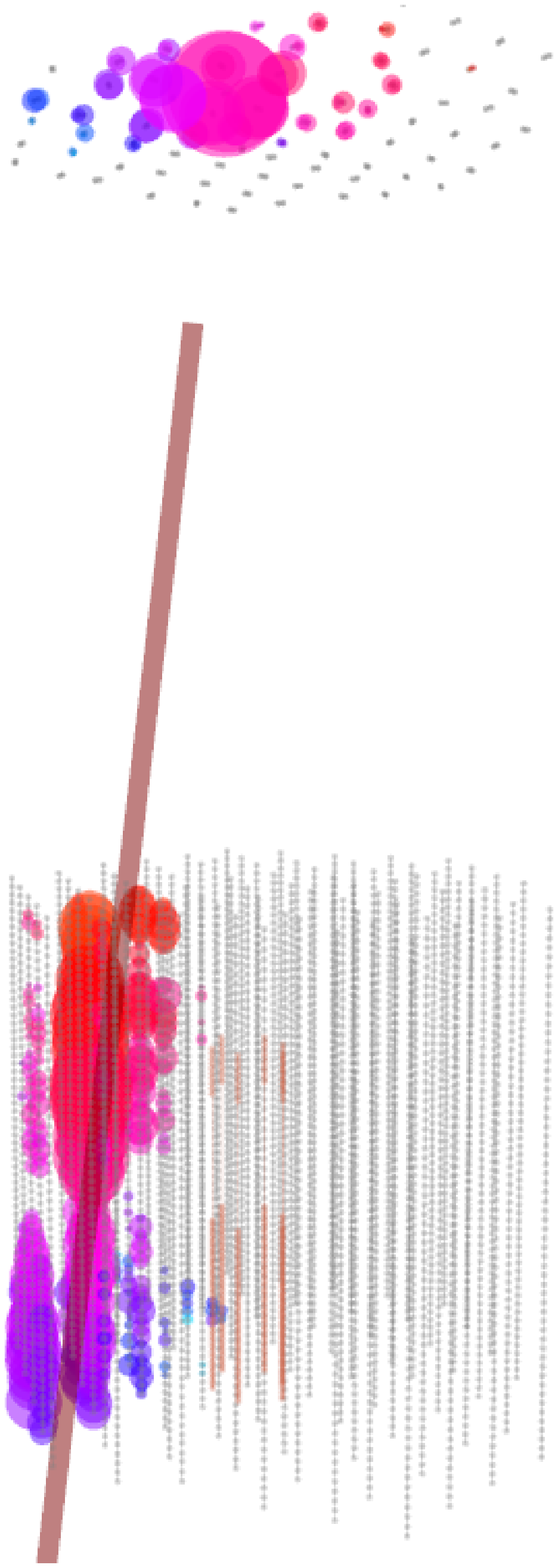}
\includegraphics*[width=6.5cm]{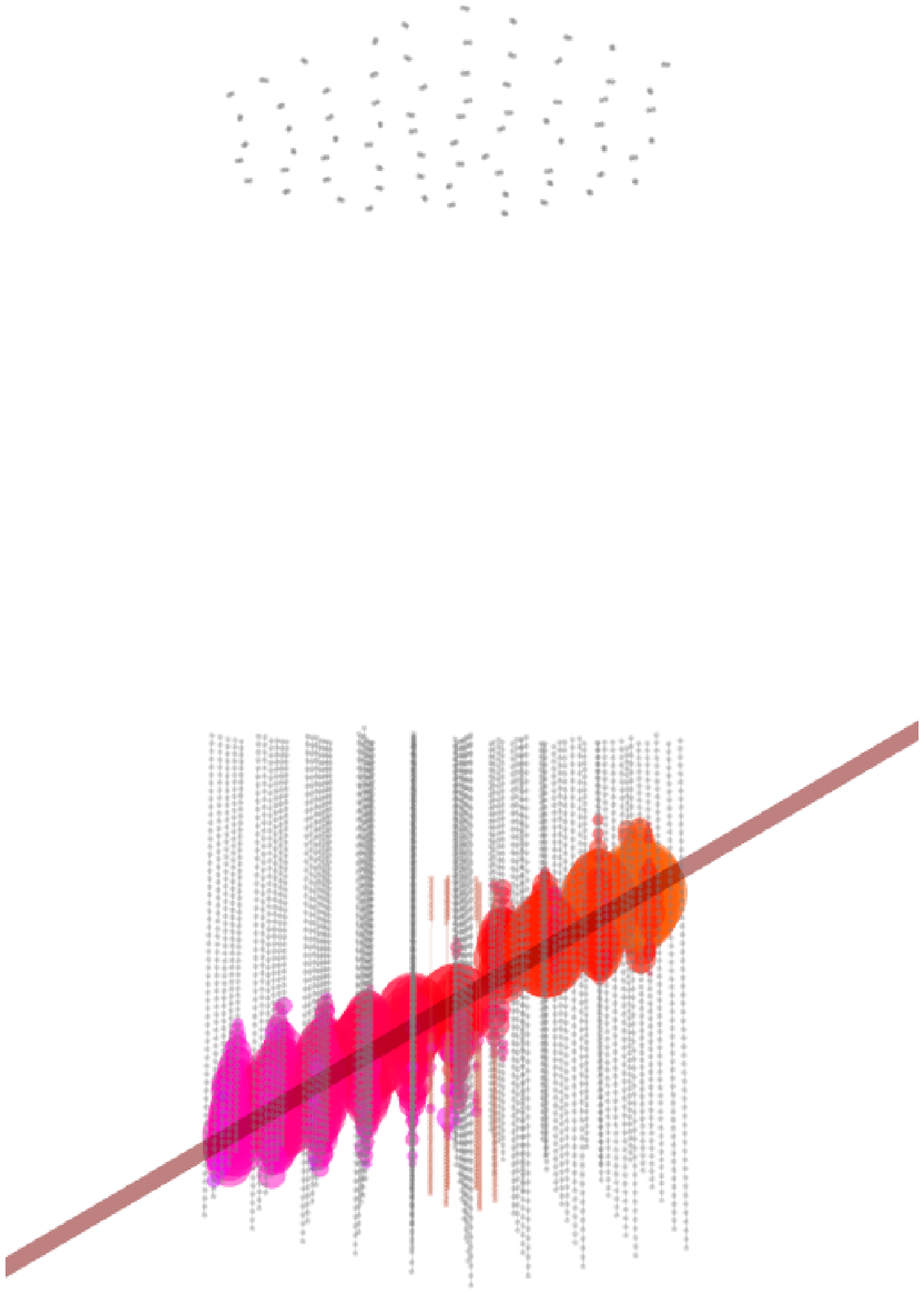}
\end{center}
\caption{Event displays: left, a nearly vertical cosmic-ray cascade
seen in coincidence by IceTop and deep IceCube running with 79 strings
in 2010; right, a muon bundle seen in the deep part of IceCube.}
\label{fig1}
\end{figure}

This paper presents IceCube as a cosmic-ray detector
with an emphasis on its potential contributions to improving knowledge of the
spectrum and composition of the primary cosmic radiation.  There are
several different and complementary cosmic-ray measurements that IceCube can make. 
First, consider IceCube as a 3-dimensional air shower detector, consisting of an array of 
1 km$^2$ on the surface and a 1 km$^3$ array between 1.45 and 2.45 kilometers
in the ice directly below the surface array.  (See Fig.~\ref{fig0}).
There are several classes of events that such a detector can measure.
Events with trajectories that pass through both parts of the detector can
be reconstructed independently by the surface array and by the in-ice array.
The ratio of the shower size on the surface to the energy deposited by
the muon bundle in the deep ice is sensitive to composition because
heavy primaries put a larger fraction of their cascade energy into muons
than protons of the same shower size.  An example of such an event is
shown in Fig.~\ref{fig1} (left).  Its energy can be estimated from the IceTop reconstruction
as $\approx 5$~PeV.  Depending on the mass
of the primary particle, this event would have 30 to 80 muons with
sufficient energy to reach 1500 m and would
deposit 5 to 15 TeV from muon energy loss in the deep detector.

Because the string spacing of 125 m in the in-ice array
is larger than the lateral extent of such
a group of high-energy muons, it is not possible to count the tracks.  However, the number
of muons is closely related to the amount of energy deposited in the detector, which
is proportional to the amount of Cherenkov light generated.  It is therefore
possible to reconstruct the energy deposited from the observed light in the detector
and hence to obtain a measure of the primary composition given the energy of
the primary cosmic ray as determined by the surface array~\citep{Feusels}.

The geometrical acceptance of IceCube for trajectories that pass
inside both the surface and the deep arrays is $A\Omega\;\approx\;0.3$~km$^2$sr,
the exact value depending on how far inside the two sub-arrays the 
trajectories are required to lie.
This corresponds to typical zenith angles of $20^\circ$.
With this acceptance, the maximum energy above which the
intensity is too low to obtain enough events for analysis is $\sim 1$~EeV.  The acceptance
can be increased by approximately a factor of ten by using all events with zenith
angles $<\,60^\circ$.  Given the steep spectrum, this would increase
the maximum useful energy to $\sim\,3$~EeV.  There are two ways to do this.
One is to use only the IceTop information.  The other is to use
muon bundles that are well reconstructed in the deep array, including those that pass
outside of IceTop, as illustrated in Fig.~\ref{fig1} (right).  

Preliminary results show a sensitivity to primary composition
through the angular dependence of the events reconstructed with
IceTop only~\citep{Klepser}.  At a given total energy per particle,
proton showers are more penetrating so they contribute relatively more
to showers of a given size at large zenith angle than heavy primaries.
A consistent interpretation of the spectra measured at different
zenith angles requires a mixed composition.  This analysis is
at a preliminary stage, using events from an early phase of construction when
there were only 26 IceTop stations, and limited to zenith angles of $\leq46^\circ$~\citep{Kislat}.
It is promising because it will be complementary to
results obtained from the ratio of high-energy muons to shower size
in coincident events.  

Another complementary approach to composition using only IceTop is possible
if information about the fraction of muons in the shower front at
the surface can be obtained.  Accomplishing this is non-trivial
because the IceTop detectors do not identify particles on an event by event
basis, but only give a measure of the energy deposited in the tank.
Efforts to measure the muon component at the surface are underway using energy deposition
in the tank to identify muons at large core distances~\citep{Hermann}.
This measurement will provide complementary information
because low-energy muons at the surface reflect later
stages of shower development as compared to high-energy muons that penetrate
to the deep detector.

The angular resolution for reconstruction 
of muons bundles in the deep part of IceCube is at the level of 1--2 degrees.
This is sufficient to determine the location of the shower core at the surface.  For events
of sufficiently high energy (depending on how far outside of IceTop the core is)
it will then be possible to reconstruct surface shower size and extend the 
geometrical acceptance for coincidence measurements with both the IceTop
and in-ice components of IceCube.

Yet another complementary approach is to measure the atmospheric muon flux
as a function of energy deposition and angle with the deep component of
IceCube only.  Preliminary results~\citep{Patrick2} with the 22 strings of IceCube operating
in 2007-08 show that the atmospheric muon spectrum can be measured with the full IceCube 
up to several hundred TeV, which probes the primary spectrum well into
the knee region of the spectrum.  One possibility is to identify single
energetic muons by measuring the characteristic large bursts of light
due to bremsstrahlung and hadronic interactions along their trajectories.

IceCube is sufficiently large to measure the spectrum of atmospheric
neutrinos into the region above 100 TeV~\citep{Warren}.  This gives a very clean
sample of events, the interpretation of which depends on assumptions about the shape
and composition of the primary spectrum in the knee region.  In the
case of both muons and neutrinos, the contribution of the prompt 
component from charm decay is also of interest.  For neutrinos there
is also the expected contribution of astrophysical neutrinos
from unresolved point sources, which is after all the main goal
of IceCube. 

Effects of composition
and charm have to be separated by their different characteristic
energy and angular dependence.  
Kaons and pions in the TeV energy
range and above have decay lengths
longer than their interaction lengths, so the intensity of neutrinos
and muons from their decays is suppressed by one power of energy
relative to the primary spectrum.  In addition the intensity is proportional
to $1/\cos{\theta}$ because the interaction length is longer
in the low-density upper atmosphere.  Charmed hadrons decay promptly because of
their short lifetimes, so above the threshold region, neutrinos and muons
from charm decay have the same spectral index as the primary spectrum,
and they are isotropic.
The astrophysical component is present
only in the neutrinos and is therefore in principle distinguishable
by comparisons between muons and neutrinos.

The power of IceCube as a cosmic-ray detector is that it can make several
different and complementary measurements that relate to the
energy-dependence of the primary composition in different ways.
They must all give a consistent result.  The energy region from PeV to EeV
is of particular interest in connection with the transition from galactic
to extra-galactic cosmic rays.

\begin{figure}
\begin{center}
\includegraphics*[width=8.0cm]{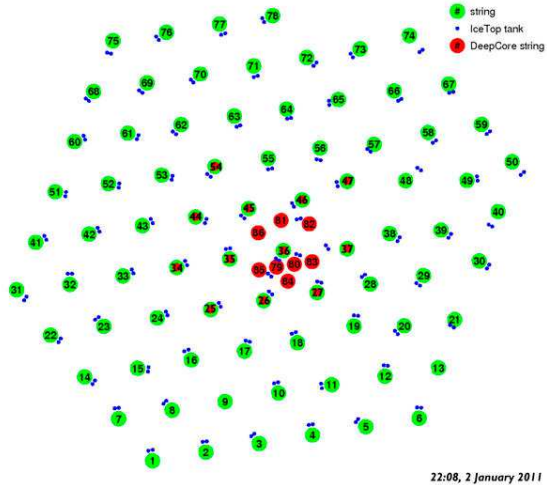}
\end{center}
\caption{Surface map of IceCube.  Numbered circles indicate location of 
IceCube strings.  Pairs of blue dots mark IceTop tanks.}
\label{surface}
\end{figure}

\section{IceTop}

IceTop in its final configuration consists of 81 stations on
the surface above the deep part of IceCube at an average pressure depth
of approximately 680 g/cm$^2$.  The layout is shown in Fig.~\ref{surface}.
Each station has two tanks separated from each other by 10 meters.  Each
pair of tanks is located 25 meters
from the point on the surface directly above
one of the deep IceCube strings.  Thus the average spacing between
neighboring IceTop stations is 125 meters, the same as for the in-ice
strings.  There are, however, significant
deviations from the regular triangular grid for the surface array
caused in some cases by nearby buildings
and in others by the contingencies of the deep drilling process which sometimes required
tanks at adjacent stations to be oriented differently with respect to their associated strings.
There are several more closely spaced stations in the
center of the array that can be
used to select a low-energy subset of events, as described in the next section.

\begin{figure}
\begin{center}
\includegraphics*[width=7.5cm]{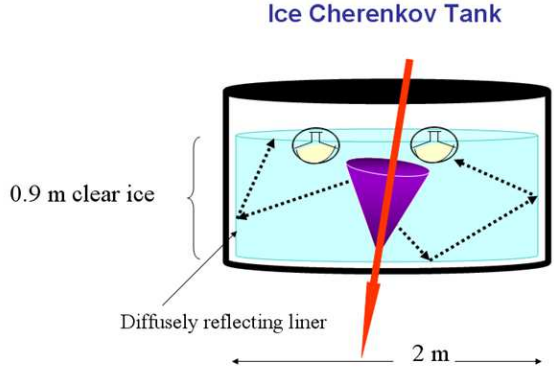}
\end{center}
\caption{
Diagram of an IceTop ice Cherenkov tank.  The diffusely
reflecting liner illuminates the photomultipliers in a way that
is well correlated with the total track length of charged particles
in the tank.  The fluctuations in the response of a single tank to
a given event are
small compared to the intrinsic fluctuations in the shower front
as measured by comparing signals from tanks at the same station in each
event~\citep{FirstYear}.}
\label{fig2}
\end{figure}

Tanks are 2 m in diameter filled with water to a depth of 90 cm.  The tanks are insulated
and instrumented during a 50 day freeze-in period with freeze-control units (FCUs) to manage the
expansion as the water freezes from the top down.  Tank tops are open while
the water freezes.  The FCUs are mounted in insulated
compartments on the outside of each tank.  In addition to managing the
expansion water, the FCUs also control a vacuum system connected
to a filter on the bottom of the tank that removes dissolved gas from the water to prevent
formation of bubbles.  When the freeze is complete, the FCUs are removed, the space above the
ice is filled with insulation and the tanks closed.

Each tank is equipped with two standard IceCube digital optical modules (DOMs) with
$10$~inch photomultipliers~\citep{PMTpaper}, 
each enclosed in a pressure sphere along with the electronics 
for digitization and time keeping.  The IceTop DOMs are fully integrated into
the IceCube data acquisition system~\citep{DAQpaper}.
Fig.~\ref{fig2} illustrates how the IceTop detector is illuminated
by Cherenkov light diffusely reflected from the white, inner lining
of the tank.  One DOM runs at low gain and the other
at high gain to give a dynamic range for each tank of more than 4 orders of magnitude.
The DOMs are mounted facing down with the
photo cathode half of the pressure sphere in the water.  Basic calibration is provided by
the spectrum of low energy cosmic-ray hits in the tanks, which occur at a rate of approximately
2 kHz per tank.  The tank spectrum consists of a low-energy electromagnetic component and
a muon peak at higher energy.  The peak is used to define a signal equivalent to a vertical muon (VEM).
As for Auger~\citep{AugerCalib}, Cherenkov light generated by air shower signals is measured in units of VEM.
For IceTop tanks, one VEM corresponds to approximately 150 to 250 photoelectrons 
in the photomultiplier, depending on which of two types of liner the tank has~\citep{Levent}.
Thus signals in IceTop tanks are significantly larger than the signals in the deep ice,
which are typically at the level of one or a few photoelectrons.

In addition to their utility for calibration, the counting rates produced in the tanks by
low energy cosmic radiation are available for heliospheric studies of solar modulation and
solar energetic particle events.  The ground-level solar particle event of December 13, 2006
was studied with data taken with the 32 tanks then in operation~\citep{solar}.

The basic air shower trigger of IceTop requires 6 DOMs to report signals within a 5$\mu$s time window.
For purposes of triggering, a DOM only reports if its neighbor is also hit, a condition called
``local coincidence".  For IceTop, the local coincidence is configured so that the neighbor of a
given DOM is always in the other tank at the same station.  Thus, even if only high-gain DOMs are
above threshold, this trigger includes
all 3-station events.  In presentations so far, however, only events involving both tanks at 5 or more stations
have been used for reconstruction and analysis.  The 5-station software trigger corresponds to an energy
threshold for full efficiency of 1 PeV for vertical showers.  The energy threshold can be lowered
by using the three (and four) station events, as discussed in the next section.


\section{Overlap with direct measurements}

As is the case with
all air shower detectors, IceCube does not measure the primary cosmic rays directly
but only samples their cascades deep in the atmosphere.  Information about
primary composition must be inferred from measurements of the properties of the
secondary air showers.  In this situation it is important to extend the indirect
measurements to sufficiently low energy to overlap as much as possible the 
energy range probed by direct measurements.  Recent measurements with ATIC~\citep{ATIC}
and CREAM~\citep{CREAM} balloon-borne calorimeters are providing increased statistics 
and a new view of
the region around 100 TeV/nucleus and above, thereby stimulating a renewed interest
in an energy region previously studied primarily with balloon-borne emulsion 
chambers, JACEE~\citep{JACEE} and RUNJOB~\citep{RUNJOB}.\footnote{See also~\citet{Kopenkin}
for a more recent analysis of RUNJOB data.}
By selecting a subset of small events in IceTop it is possible to obtain an overlap in energy
with direct measurements down to $<\,300$~TeV~\citep{Bakhtiyar}.

\begin{figure}[thb]
\begin{center}
\includegraphics*[width=6.5cm,angle=-90]{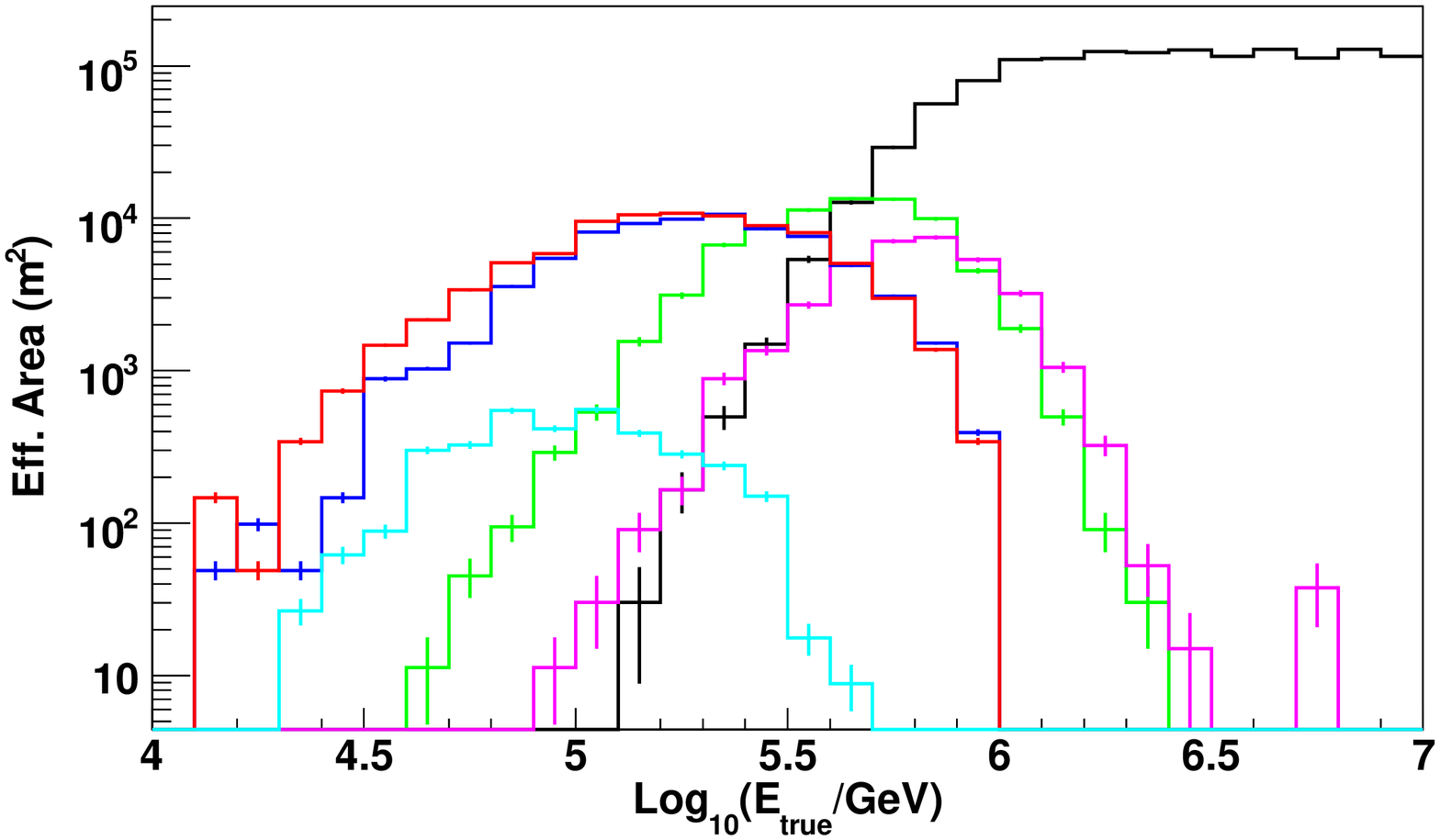}
\includegraphics*[width=6.5cm,angle=-90]{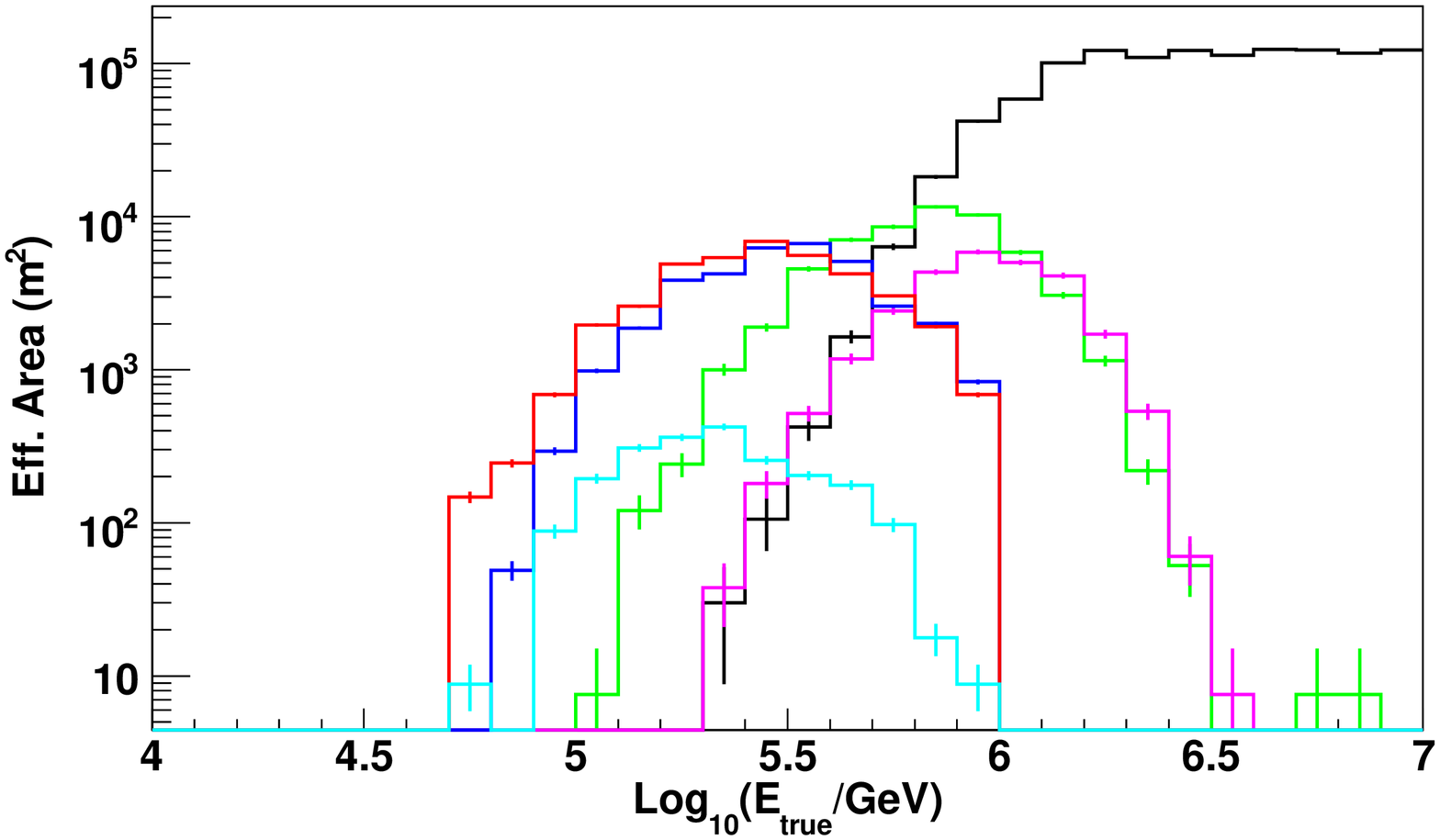}
\end{center}
\caption{Effective area as a function of energy for various 
classes of events.  Left panel is for primary protons, the right panel
for primary iron.  The black line is for the standard IceTop analysis
which requires 5 or more stations to have hits in both tanks, which
becomes fully efficient for vertical events above a PeV.  For IceTop 40
shown here, the cores were required to be located in the interior
portion of the array with an area of $0.11$~km$^2$.  Each plot has two
sets of colored histograms: the green histograms are for 3-station events
in the main IceTop array, while the pink histograms at slightly higher energy
are for 4-station events.  The histograms at lower energy were made
for planning the locations of more closely spaced stations in the center
of the array.}
\label{fig4}
\end{figure}

Fig.~\ref{fig4} shows the effective area obtained from simulations separately for
protons and for iron primaries for 3- and 4-station events
compared to the acceptance for IceTop when it was running with 40 stations.  The acceptance curves
for 3- and 4-station events at first
rise as energy increases and then decrease as more stations are hit.
Thus, for example, the sum of all 3-station events in the interior of
IceTop-40 peaks at 300 TeV for primary protons and at just below a PeV
for primary iron.  The three station trigger therefore preferentially selects
light primaries.  Selecting a particular group of stations with smaller
spacing, it is possible to shift the response to somewhat lower energy.
Studies of the final array configuration with the goal of overlapping 
direct measurements are currently underway.

\section{Muons in IceCube}
The amount of data on TeV muons being accumulated by IceCube is enormous.
At a rate of more than 2 kHz, the deep IceCube detector collects
almost 100 billion events per year.  In addition to the potential for
constraining the energy dependence of the cosmic-ray composition in
the knee region, the high rate of muons in IceCube also supports
detailed studies of cosmic-ray anisotropy~\citep{Simona}.  The IceCube
map extending cosmic-ray anisotropy studies into the Southern hemisphere
is shown by~\citet{Ty} in this volume.  

Another aspect of the high rate of cosmic-ray muons in IceCube is that
the characteristic correlation of atmospheric muons at high energy with
the temperature in the stratosphere can be used as an additional probe
of the relative contributions of pions, kaons and charmed hadrons to
the flux of leptons in the atmosphere~\citep{DesiatiGaisser}.  The variation
observed with the 40-string version of IceCube is reported by~\citet{Tilav}.

\noindent
{\bf Acknowledgment} This research is supported in part by the U.S.
National Science Foundation under Grant 0856253.

\end{document}